\documentclass[a4paper,12pt]{article}
\usepackage{tabularx}
\usepackage{array}
\usepackage[
a4paper,
top=1.5cm,
bottom=2cm,
left=2.5cm,
right=2.5cm
]{geometry}
 \usepackage{tikz}
\usetikzlibrary{arrows.meta,positioning,calc,shapes.geometric}
\usepackage[utf8]{inputenc}
\usepackage{amsmath}
\usepackage{graphicx}
\usepackage{booktabs}
\usepackage{longtable}
\usepackage{array}
\usepackage{multirow}
\usepackage{adjustbox}
\usepackage{float}
\usepackage{pdflscape}
\usepackage{xcolor}
\usepackage{listings}

\usepackage{wrapfig}

\usepackage{subcaption}

\usepackage{tikz}
\usetikzlibrary{positioning,arrows.meta}

\usepackage{setspace}
\usepackage{hyperref}
\usepackage{cleveref}
\usepackage{authblk}

\usepackage{newunicodechar}
\newunicodechar{⁻}{\textsuperscript{-}}

\usepackage{apacite}
\usetikzlibrary{positioning}

\title{
\Large
AI-driven Multimodal Representation Learning for Latent Mediation Structure Discovery of Socioeconomic Disadvantage, Psychosocial Factors, and Cardiometabolic Multimorbidity: Insights from the All of Us Research Program
}

\author[1]{Cong Cao}
\author[1]{Shuangge Ma\thanks{Corresponding author: Shuangge Ma, Department of Biostatistics, Yale School of Public Health, Yale University, New Haven, CT 06520, USA. Email: \texttt{shuangge.ma@yale.edu}}}

\affil[1]{Department of Biostatistics, Yale School of Public Health, Yale University, New Haven, CT 06520, USA}

\date{} 

\date{\vspace{-3em}}
\begin{document}
\maketitle
\vspace{-2em}

\section*{Abstract}

Social disadvantage is associated with multimorbidity, but the pathways linking social conditions to disease burden remain poorly understood. We developed an AI-driven multimodal mediation framework that integrates socioeconomic, psychosocial, clinical, laboratory, behavioral, and genomic data from the All of Us Research Program. Modality-specific variational autoencoders were used to derive latent representations of each data domain, and mediation analyses were subsequently performed in latent space to evaluate indirect associations between socioeconomic disadvantage, psychosocial factors, and multimorbidity. The final analytic cohort included 20,804 participants with complete multimodal data. Across 800 exposure--mediator--outcome combinations, mediation signals were concentrated within a small number of latent dimensions. The strongest indirect association linked a socioeconomic disadvantage dimension, a psychosocial vulnerability dimension, and a cardiometabolic multimorbidity dimension (NIE = 0.002517). The psychosocial dimension was characterized by poorer mental health, greater loneliness, lower social well-being, and lower health literacy, whereas the outcome dimension was associated with hypertension, diabetes, hyperlipidemia, obesity, chronic kidney disease, and heart disease. Bootstrap analyses supported the stability of the leading pathway. These findings suggest that psychosocial vulnerability was strongly represented in the dominant latent pathway linking socioeconomic disadvantage and cardiometabolic multimorbidity. More broadly, the proposed framework illustrates how AI-based representation learning can be used to investigate complex relationships across high-dimensional multimodal health data.

\textbf{Keywords:}
Multimorbidity; Social Determinants of Health; Mediation Analysis; Artificial Intelligence; Variational Autoencoder; All of Us Research Program

\section*{Introduction}
 
Health is shaped by more than biology. Income, education, housing conditions, and other social determinants influence the risk of chronic disease, multimorbidity, and premature mortality. Although these associations are well established, the mechanisms through which social disadvantage becomes linked to adverse health outcomes remain incompletely understood. Psychosocial factors, including mental health, loneliness, social support, and health literacy, have been proposed as important intermediaries linking socioeconomic conditions to cardiovascular and metabolic diseases \cite{toutounji2024stress,delmas2025linking}. However, most previous studies have focused on a limited number of psychosocial factors and individual disease outcomes, making it difficult to characterize the complex pathways through which social experiences may influence multimorbidity across multiple health domains.

Mediation analysis provides a useful framework for examining potential pathways linking exposures, intermediate variables, and health outcomes \cite{baron1986moderator,Zhao2022}. Classical mediation approaches typically rely on parametric regression models and often assume relatively simple relationships among exposures, mediators, and outcomes \cite{byun2023exploring}. These assumptions become increasingly restrictive in modern population cohorts, where variables are high-dimensional, highly correlated, and potentially related through nonlinear interactions \cite{verma2024mediation}. As a result, identifying pathways within complex biomedical systems remains challenging using conventional mediation frameworks alone.

Recent methodological developments have increasingly incorporated machine learning into mediation and pathway analyses. Neural-network–based approaches have been used to model complex mediator-outcome relationships and accommodate nonlinear dependencies that are difficult to capture using traditional regression models \cite{ali2022soft,edwards2021associations,xu2022deepmed}. Other studies have developed machine-learning methods for high-dimensional mediation settings involving large numbers of variables and complex dependence structures \cite{Pagare2024,zhou2025causal,jiang2023causal}. In parallel, representation-learning techniques have been applied to behavioral, clinical, and biological data to derive lower-dimensional latent features suitable for downstream analyses \cite{dong2024personality,siegersma2022deep}, while generative models have been explored for estimating counterfactual quantities in more complex settings \cite{lu2025influence}. Together, these developments have expanded opportunities for investigating pathways within large-scale multimodal health datasets.

Multimorbidity presents a particularly challenging setting because chronic diseases occur in interconnected patterns rather than in isolation. Disease outcomes are often correlated through shared biological, behavioral, and social determinants, creating network-like structures that are difficult to represent using conventional analytical approaches \cite{kojima2022brain,Zhang2025}. Despite growing interest in representation learning and high-dimensional mediation methods, relatively few studies have combined multimodal representation learning, psychosocial pathways, and multimorbidity outcomes within a unified population-health framework.

The All of Us Research Program provides linked survey, electronic health record, laboratory, behavioral, and genomic data within a single nationwide cohort. These data allow social, psychosocial, clinical, and biological domains to be examined jointly, but their high dimensionality and complex correlation structure present substantial analytical challenges. Many variables are correlated within and across modalities, and multimorbidity outcomes are naturally organized through disease co-occurrence rather than isolated diagnoses.

To address these challenges, we combined multimodal representation learning with latent-space pathway analysis in the All of Us Research Program. Modality-specific variational autoencoders were used to derive low-dimensional representations of socioeconomic, psychosocial, disease, laboratory, behavioral, and genomic domains, which were subsequently integrated within a unified analytical framework. Our objective was to investigate potential pathways linking socioeconomic disadvantage, psychosocial vulnerability, and cardiometabolic multimorbidity in a large multimodal population cohort \cite{jiang2023causal,siegersma2022deep,dong2024personality}.

Throughout this study, mediation terminology is used to describe statistical decomposition of indirect associations rather than identification of causal mediation effects. We use the terms natural indirect effect (NIE) and natural direct effect (NDE) as convenient labels for regression-based decomposition quantities computed in latent space. These quantities do not have a counterfactual causal interpretation and should not be interpreted as causally identified mediation effects.

\section*{Data Sources and Study Population}
Data were obtained from the All of Us Controlled Tier v8 data release, and it included participant-level information from questionnaires, electronic health records, laboratory measurements, behavioral assessments, and whole-genome sequencing data. These data were organized into seven analytic domains: socioeconomic exposures ($T$), psychosocial mediators ($M$), disease outcomes ($Y$), demographic covariates ($X$), laboratory covariates ($L$), behavioral covariates ($C$), and genomic burden profiles ($G$).

Socioeconomic exposure variables were derived from The Basics and Social Determinants of Health survey instruments. All combinations of exposure, mediator, and outcome dimensions were evaluated, resulting in 800 candidate mediation pathways. Baseline demographic covariates included sex at birth and race/ethnicity. These variables were used as adjustment factors in downstream analyses. Psychosocial variables were derived from self-reported survey measures and were available for 633,537 participants. Measures included general mental health, emotional distress, loneliness-related indicators, social well-being, social functioning, social support, and health-literacy items. These variables were considered candidate mediators because they capture psychological, social, and informational dimensions that may connect socioeconomic conditions to downstream health outcomes.

Disease outcomes were derived from electronic health records. Binary disease indicators were constructed to represent the presence or absence of chronic conditions. Disease-outcome data were available for 256,530 participants. The disease list included clinically defined conditions spanning major organ systems, including cardiovascular, metabolic, respiratory, mental health, musculoskeletal, digestive, renal, neurologic, ophthalmologic, dermatologic, and infectious disease categories. Disease co-occurrence relationships were subsequently used to construct disease-network phenotypes for downstream modeling.

 Clinical covariates included laboratory and vital-sign measurements available for 474,829 participants. Variables included body mass index, systolic blood pressure, diastolic blood pressure, glucose, glycated hemoglobin (HbA1c), creatinine, total cholesterol, LDL cholesterol, HDL cholesterol, and triglycerides. Behavioral variables were derived from participant survey responses and were available for 215,221 participants. Variables included smoking history, smoking frequency, alcohol participation, alcohol frequency, and alcohol quantity. Whole-genome sequencing data were used to construct rare-variant burden profiles. Sequencing data were available for 414,830 participants. Rare variants were aggregated into gene-level burden features to generate genomic burden matrices for downstream representation learning. After burden construction and feature-availability filtering, the genomic analytic dataset included 103,867 participants and 310 gene-level burden features.

Participant-level data from all modalities were linked using unique participant identifiers. The final mediation analysis retained participants with complete information across socioeconomic exposure, psychosocial mediator, disease outcome, demographic, laboratory, behavioral, and genomic domains. After complete-case integration, the final analytic cohort consisted of 20,804 participants.

 \section*{Methods}

\begin{figure}[htbp]
\centering
\includegraphics[width=\textwidth]{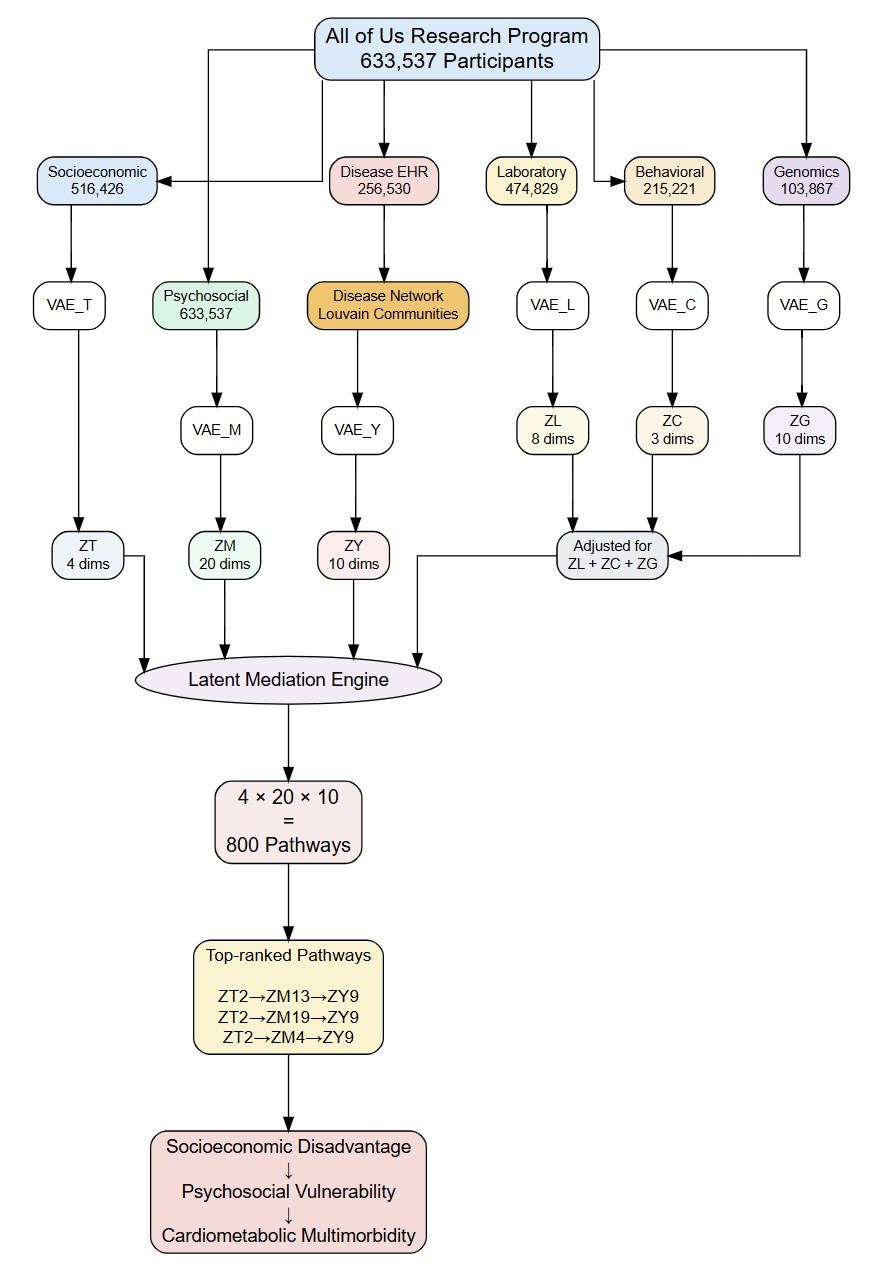}
\caption{
Overview of the AI-driven multimodal mediation framework. Disease co-occurrence networks were first constructed from electronic health records. Separate variational autoencoders were then trained for socioeconomic, psychosocial, disease, laboratory, behavioral, and genomic data to derive latent representations. Participant-level latent embeddings were subsequently integrated and used in latent-space mediation analyses to evaluate pathways linking socioeconomic disadvantage, psychosocial vulnerability, and cardiometabolic multimorbidity while adjusting for demographic, laboratory, behavioral, and genomic factors.
}
\label{fig:workflow}
\end{figure}

We developed a multimodal mediation framework to investigate pathways linking socioeconomic conditions, psychosocial factors, and multimorbidity in the All of Us Research Program. Because each data modality contained numerous correlated variables, low-dimensional latent representations were learned before mediation analysis. This framework enabled the evaluation of indirect pathways connecting socioeconomic exposures, psychosocial factors, and disease outcomes while accounting for demographic, behavioral, clinical, and genetic information. An overview of the analytical workflow is shown in Figure~\ref{fig:workflow}.

The workflow consisted of four stages. First, disease co-occurrence networks were constructed from electronic health record data to characterize patterns of multimorbidity. Second, separate variational autoencoders (VAEs) were trained for each data modality to learn participant-level latent representations. Independent models were used because modalities differed substantially in scale, sparsity, distribution, and clinical meaning. Third, latent representations from all modalities were linked at the participant level using unique participant identifiers. Finally, mediation analyses were performed to evaluate pathways linking socioeconomic exposures, psychosocial factors, and disease outcomes while adjusting for demographic, laboratory, behavioral, and genomic factors.

Disease co-occurrence networks were constructed using diagnoses observed in the All of Us cohort. Nodes represented disease categories and edges represented pairwise disease co-occurrence relationships. Network structure was characterized using the Louvain community detection algorithm, and modularity was used to quantify the strength of community organization. To further evaluate multimorbidity structure, network analyses were performed in both the full cohort and a high-multimorbidity subgroup comprising individuals with multiple chronic conditions. Network-derived disease features were encoded at the participant level and used as inputs to the disease-specific VAE.

\subsection*{Model Training and Latent Representation Extraction}

For each data modality, a separate VAE was trained to learn a low-dimensional participant-level representation. Before model fitting, continuous variables were standardized, categorical variables were encoded as model input features, and rare-variant burden counts were transformed using a log$(1+x)$ transformation before standardization. For the genomic module, rare variants were aggregated into 310 gene-level burden features, producing the input matrix for the genomic VAE.

Each VAE consisted of an encoder, a latent layer, and a decoder. The encoder mapped modality-specific input features to a lower-dimensional latent distribution, and the decoder reconstructed the original inputs from sampled latent representations. Models were trained by minimizing the standard VAE objective, combining reconstruction loss and a Kullback--Leibler divergence penalty. Posterior encoder means were extracted as participant-level latent embeddings for downstream analyses.

Model tuning used an 80\% training and 20\% validation split. Candidate latent dimensions, hidden-layer widths, learning rates, and KL-divergence penalty weights were evaluated using validation-set evidence lower bound (ELBO) loss. The search considered latent dimensions of 5, 10, 20, and 50; hidden-layer widths of 128 and 256 units; learning rates of $10^{-3}$ and $10^{-4}$; and KL-divergence penalty weights of 0.001 and 0.01. The final configuration was selected based on the lowest validation ELBO loss. The final latent dimensionality for each modality was selected independently according to validation ELBO performance.

Models were implemented in PyTorch and trained using the Adam optimizer for 100 epochs with a mini-batch size of 1,024. The resulting latent spaces contained four socioeconomic dimensions ($Z_T$), twenty psychosocial dimensions ($Z_M$), ten disease-network dimensions ($Z_Y$), ten genomic dimensions ($Z_G$), eight laboratory dimensions ($Z_L$), and three behavioral dimensions ($Z_C$). These latent representations were then used in downstream mediation analyses. Detailed architecture and hyperparameter settings are provided in the Supplementary Methods.

\subsection*{Latent-Space Mediation Estimation}

The mediation analysis evaluated whether psychosocial factors may represent intermediate latent dimensions associated with the linkage between socioeconomic disadvantage and multimorbidity. Socioeconomic latent dimensions were treated as exposures ($Z_T$), psychosocial latent dimensions as mediators ($Z_M$), and disease-network latent dimensions as outcomes ($Z_Y$), yielding pathways of the form

\[
Z_T \rightarrow Z_M \rightarrow Z_Y.
\]

For each exposure dimension $Z_{Tj}$, mediator dimension $Z_{Mk}$, and outcome dimension $Z_{Yl}$, two regression models were fitted. The mediator model was

\[
Z_{Mk} = \alpha_0 + a_{jk}Z_{Tj} + \gamma^\top W + \epsilon_M,
\]

where $a_{jk}$ denotes the exposure--mediator association and $W$ includes demographic, laboratory, behavioral, and genomic latent representations.

The outcome model was

\[
Z_{Yl} = \beta_0 + c'_{jl}Z_{Tj} + b_{kl}Z_{Mk} + \delta^\top W + \epsilon_Y,
\]

where $b_{kl}$ denotes the mediator--outcome association after adjustment and $c'_{jl}$ denotes the direct exposure--outcome association.

Natural indirect effects (NIEs) were estimated using the product-of-coefficients approach:

\[
\mathrm{NIE}_{jkl} = a_{jk} \times b_{kl}.
\]

Because all variables were represented as latent dimensions, indirect effects were interpreted as standardized pathway strengths in latent space rather than direct changes in disease incidence or clinical risk.

The exposure representation contained four dimensions, the psychosocial representation contained twenty dimensions, and the disease representation contained ten dimensions, yielding

\[
4 \times 20 \times 10 = 800
\]

candidate mediation pathways. All models were adjusted for demographic ($Z_X$), laboratory ($Z_L$), behavioral ($Z_C$), and genomic ($Z_G$) embeddings. Pathways were ranked according to the absolute magnitude of NIEs, while signs were retained for interpretation. False discovery rate correction was applied to account for multiple testing, and the strongest pathway was further evaluated using nonparametric bootstrap resampling.

  \section*{Results}
\subsection*{Cohort Description}

 The analytic cohort was drawn from participants in the All of Us Research Program with available socioeconomic, psychosocial, disease, laboratory, behavioral, and genomic data. Data availability varied considerably across data sources. Psychosocial survey data were available for 633,537 participants, socioeconomic exposure data for 516,426 participants, laboratory measurements for 474,829 participants, electronic health record–derived disease outcomes for 256,530 participants, and behavioral survey data for 215,221 participants. Whole-genome sequencing data were available for 414,830 participants. After constructing rare-variant burden measures and applying quality-control filters, the genomic dataset included 103,867 participants. To ensure complete information across all exposure, mediator, outcome, and covariate domains, only participants with data available for all required modalities were included in the mediation analyses. After integrating data across all sources, the final analytic cohort consisted of 20,804 individuals. The reduction in sample size was primarily due to incomplete overlap among behavioral, disease, and genomic data sources, which limited the number of participants with complete information across all analytic domains (see \cref{tab:data_modalities,tab:network_summary}).
 
\begin{table}[htbp]
\centering
\caption{Participant counts and analytic role of each data modality}
\label{tab:data_modalities}
\small
\begin{tabular}{llcc}
\hline
Modality & Analytic role & Participants & Latent dimensions \\
\hline
Socioeconomic survey data & Exposure ($Z_T$) & 516,426 & 4 \\
Psychosocial survey data & Mediator ($Z_M$) & 633,537 & 20 \\
Electronic health records & Outcome ($Z_Y$) & 256,530 & 10 \\
Demographic survey data & Covariate ($Z_X$) & 516,426 & 2 \\
Laboratory measurements & Covariate ($Z_L$) & 474,829 & 8 \\
Behavioral survey data & Covariate ($Z_C$) & 215,221 & 3 \\
Whole-genome sequencing available & Genomic source data & 414,830 & -- \\
Gene-burden matrix after filtering & Covariate ($Z_G$) & 103,867 & 10 \\
Final integrated cohort & Mediation analysis & 20,804 & -- \\
\hline
\end{tabular}
\end{table}

\begin{table}[htbp]
\centering
\caption{Characteristics of disease co-occurrence networks in the All of Us cohort.}
\label{tab:network_summary}
\begin{tabular}{lcc}
\hline
Metric & Full Cohort & High-Multimorbidity Cohort\\
\hline
Patients retained & 257,097 & 42,424\\
Mean disease count & 1.78 & 5.68\\
Median disease count & 1.0 & 5.0\\
Maximum disease count & 24 & 22\\
Nodes & 28 & 22\\
Edges & 316 & 24\\
Communities & 3 & 6\\
Modularity & 0.116 & 0.694\\
\hline
\end{tabular}
\end{table}

   Disease co-occurrence networks showed markedly different patterns in the full cohort and the high-multimorbidity subgroup (Table~\ref{tab:network_summary}). The full cohort network included 257,097 participants, 28 disease nodes, and 316 connections between diseases. Community detection identified three major disease clusters, with a modularity score of 0.116. To better understand multimorbidity patterns, we also examined a subgroup of individuals with multiple co-occurring conditions. This high-multimorbidity network included 42,424 participants, 22 disease nodes, and 24 connections. Community detection identified six disease clusters and a much higher modularity score (0.694), indicating stronger separation between disease communities than in the full cohort. Participants in the high-multimorbidity subgroup had an average of 5.68 conditions per person (median = 5), compared with 1.78 conditions per person in the full cohort.

 \subsection*{Mediation Results}
Although indirect effects were small in absolute magnitude, this was expected because mediation analyses were performed using standardized latent representations. Accordingly, interpretation focused on the relative importance of pathways rather than coefficient magnitude alone. Among exposure dimensions, $Z_{T2}$ exhibited the strongest overall mediation activity, with a mean absolute indirect effect of $5.51\times10^{-4}$, approximately 3.3 times larger than that of $Z_{T1}$ ($1.65\times10^{-4}$). Among mediator dimensions, $Z_{M13}$, $Z_{M19}$, and $Z_{M4}$ consistently generated the largest indirect effects, whereas mediation signals were strongest for outcome dimension $Z_{Y9}$, followed by $Z_{Y4}$ and $Z_{Y8}$. Overall, the mediation structure was driven by a relatively small subset of latent dimensions (Supplementary Table~\ref{tab:mediation_summary}). Complete rankings of exposure, mediator, and outcome dimensions are provided in Supplementary Table~\ref{tab:mediation_summary}. Figure~\ref{fig:exposure_outcome} summarizes the average mediation strength across exposure--outcome dimension pairs and shows that $Z_{T2}$ produced the strongest mediation signals across multiple outcome dimensions, particularly $Z_{Y9}$.

\begin{figure}[ht]
\centering
\includegraphics[width=0.85\textwidth]{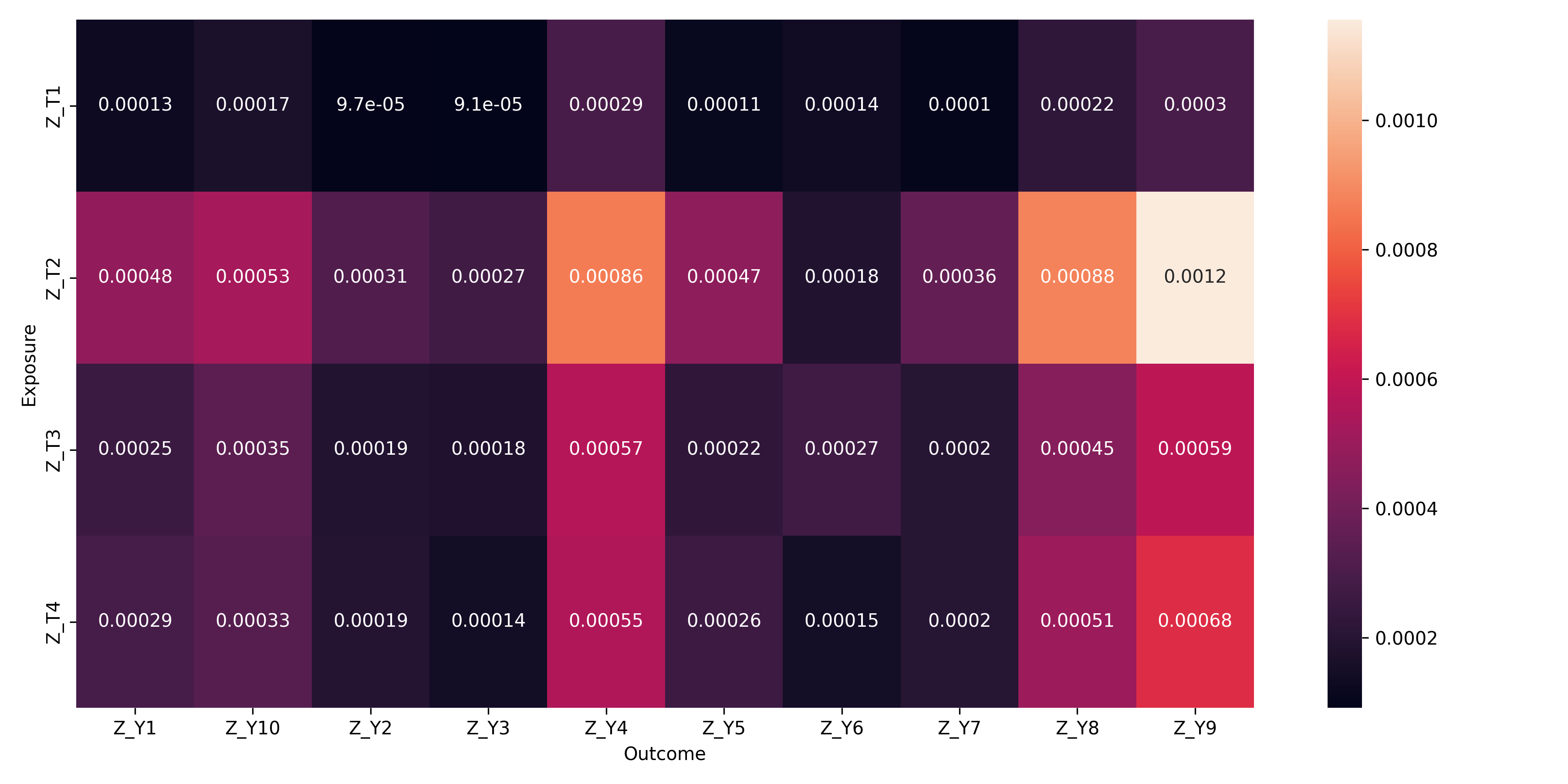}
\caption{
Mean absolute natural indirect effects (NIEs) aggregated across mediators for each exposure--outcome pair.
Darker colors indicate weaker mediation effects whereas lighter colors indicate stronger mediation effects.
}
\label{fig:exposure_outcome}
\end{figure}

 Among the 800 exposure–mediator–outcome combinations examined, indirect effect estimates varied considerably across latent dimensions. Among the mediator dimensions, $Z_{M13}$ showed the largest average absolute indirect effect, followed by $Z_{M19}$, $Z_{M4}$, $Z_{M5}$, and $Z_{M12}$. Among the outcome dimensions, $Z_{Y9}$ had the largest average absolute indirect effect, followed by $Z_{Y4}$ and $Z_{Y8}$. The strongest individual indirect association was observed for the pathway

\[
Z_{T2} \rightarrow Z_{M13} \rightarrow Z_{Y9},
\]

with an estimated latent-space NIE of 0.002517.

 Several mediator dimensions showed different mediation patterns depending on the outcome dimension. For example, $Z_{M13}$ was associated with positive indirect effects for $Z_{Y9}$ but negative indirect effects for $Z_{Y8}$ and $Z_{Y4}$. Significant indirect effects were observed for both positive and negative pathways, indicating that the direction of mediation varied across outcome domains. The strongest indirect associations were concentrated in a relatively small number of exposure, mediator, and outcome dimensions (Table~\ref{tab:mediation_coefficients}; Appendix, Figure~\ref{fig:top_pathways}). All pathways shown remained significant after correcting for multiple comparisons. False discovery rate was controlled using the Benjamini–Hochberg procedure. The dominant outcome dimension was primarily associated with cardiometabolic multimorbidity, including hypertension, diabetes, hyperlipidemia, obesity, chronic kidney disease, heart disease, sleep apnea, and heart failure. Although interpreted primarily as a psychosocial dimension based on its dominant survey loadings, $Z_{M13}$ also showed associations with several chronic disease indicators, suggesting that latent representations may capture shared variation across psychosocial and health domains.

 \begin{table}[htbp]
\centering
\caption{
Top-ranked latent mediation pathways.
}
\label{tab:mediation_coefficients}

\small

\begin{tabular}{lllllll}
\hline
Exposure & Mediator & Outcome & $a$ & $b$ & NIE & NDE \\
\hline

$Z_{T2}$ & $Z_{M13}$ & $Z_{Y9}$ &
0.198 & 0.0127 & 0.00252 & -0.01215 \\

$Z_{T2}$ & $Z_{M4}$ & $Z_{Y9}$ &
-0.270 & -0.00884 & 0.00239 & -0.01202 \\

$Z_{T2}$ & $Z_{M19}$ & $Z_{Y9}$ &
-0.180 & -0.01195 & 0.00215 & -0.01178 \\

$Z_{T2}$ & $Z_{M5}$ & $Z_{Y9}$ &
-0.223 & -0.00918 & 0.00205 & -0.01168 \\

$Z_{T2}$ & $Z_{M13}$ & $Z_{Y8}$ &
0.198 & -0.01003 & -0.00199 & 0.01187 \\

$Z_{T2}$ & $Z_{M12}$ & $Z_{Y9}$ &
0.345 & 0.00574 & 0.00198 & -0.01161 \\

$Z_{T2}$ & $Z_{M19}$ & $Z_{Y8}$ &
-0.180 & 0.01046 & -0.00188 & 0.01176 \\

$Z_{T2}$ & $Z_{M6}$ & $Z_{Y9}$ &
0.220 & 0.00856 & 0.00188 & -0.01151 \\

$Z_{T2}$ & $Z_{M4}$ & $Z_{Y4}$ &
-0.270 & 0.00670 & -0.00181 & 0.00409 \\

$Z_{T2}$ & $Z_{M13}$ & $Z_{Y4}$ &
0.198 & -0.00889 & -0.00176 & 0.00404 \\

\hline
\end{tabular}

\vspace{0.2cm}

\footnotesize
All displayed pathways remained significant after correction for multiple comparisons. Indirect effects were estimated using standardized latent representations, allowing effect sizes to be compared across different exposure, mediator, and outcome dimensions.

\end{table}

\vspace{0.3cm}

\noindent

 \subsection*{Bootstrap Analysis}

 Bootstrap analysis of the strongest pathway ($Z_{T2}\rightarrow Z_{M13}\rightarrow Z_{Y9}$) yielded an NIE of 0.00291 (95\% CI: 0.00243–0.00344), indicating a robust indirect association. Examination of the latent loadings showed that $Z_{T2}$ was mainly driven by socioeconomic factors. Higher levels of education and household income were associated with lower latent scores, while individuals with lower income and lower educational attainment had higher scores. This pattern suggests that $Z_{T2}$ represents socioeconomic disadvantage. Income and education were strongly negatively associated with the latent factor, whereas poorer mental health was positively associated with higher $Z_{T2}$ scores. Bootstrap analyses were restricted to the highest-ranked pathway because the objective was to assess stability of the dominant mediation signal rather than to re-evaluate all 800 pathways.

Among mediator dimensions, $Z_{M13}$ was interpreted as a latent psychosocial vulnerability dimension based on its strongest observed associations with mental health, loneliness, social well-being, and health literacy measures. This latent dimension was also positively associated with hypertension, diabetes, and hyperlipidemia, suggesting enrichment for cardiometabolic vulnerability.
To facilitate interpretation of the dominant mediation pathway, we examined the variables most strongly associated with the key latent dimensions. Higher values of $Z_{M13}$ were also associated with hypertension, diabetes, and hyperlipidemia, indicating enrichment for cardiometabolic risk. 

 To further characterize its clinical relevance, correlations between $Z_{M13}$ and major chronic diseases were examined (Table~\ref{tab:m13_interpretation}).
To further characterize the dominant outcome dimension, we examined correlations between $Z_{Y9}$ and routine clinical biomarkers. Systolic blood pressure showed the strongest association ($r=0.117$), whereas correlations with HbA1c, diastolic blood pressure, and lipid measures were comparatively modest. These findings support the interpretation of $Z_{Y9}$ as a cardiometabolic disease dimension. The leading mediation pathways were concentrated among a limited number of latent dimensions. Most of the largest indirect associations originated from $Z_{T2}$ and involved outcome dimensions $Z_{Y9}$, $Z_{Y8}$, or $Z_{Y4}$. Several mediator dimensions, including $Z_{M13}$, $Z_{M19}$, and $Z_{M4}$, appeared repeatedly among the top-ranked pathways. This pattern indicates that the strongest indirect associations were not evenly distributed across the latent space, but clustered within a small subset of exposure, mediator, and outcome dimensions.

\begin{table}[htbp]
\centering
\caption{
Top disease correlates of latent mediator Z\_M13 (cluster 3).
Correlations were calculated between latent mediator scores and disease indicators.
}
\label{tab:m13_interpretation}
\begin{tabular}{lcc}
\hline
Disease & Correlation & Prevalence \\
\hline
Hypertension & 0.385 & 24.4\% \\
Diabetes & 0.337 & 15.8\% \\
Hyperlipidemia & 0.312 & 18.2\% \\
Heart disease & 0.214 & 6.5\% \\
Chronic kidney disease & 0.208 & 6.1\% \\
Obesity & 0.206 & 9.7\% \\
Sleep apnea & 0.196 & 7.0\% \\
Heart failure & 0.182 & 4.3\% \\
Anemia & 0.171 & 7.2\% \\
Atrial fibrillation & 0.160 & 3.7\% \\
Osteoarthritis & 0.148 & 9.2\% \\
GERD & 0.144 & 8.9\% \\
\hline
\end{tabular}
\end{table}

 \section*{Discussion}

Using multimodal data from the All of Us Research Program, we identified a latent pathway linking socioeconomic disadvantage, psychosocial vulnerability, and cardiometabolic multimorbidity. Although hundreds of candidate exposure--mediator--outcome combinations were evaluated, the strongest indirect associations were concentrated within a small number of latent dimensions. In particular, a psychosocial dimension characterized by poorer mental health, loneliness, lower social well-being, and reduced health literacy was consistently linked to a cardiometabolic multimorbidity dimension encompassing hypertension, diabetes, hyperlipidemia, obesity, chronic kidney disease, and cardiovascular disease. These findings suggest that psychosocial vulnerability represents an important component of the association between socioeconomic disadvantage and cardiometabolic disease burden.

Our findings are broadly consistent with previous research demonstrating that socioeconomic disadvantage is associated with both psychosocial adversity and increased cardiometabolic risk \cite{toutounji2024stress,delmas2025linking}. Individuals facing economic and educational disadvantage are more likely to experience chronic stress, social isolation, barriers to healthcare access, and reduced health literacy, all of which have been linked to adverse cardiovascular and metabolic outcomes. Rather than examining these factors separately, our analysis suggests that they cluster within a common latent psychosocial dimension that is strongly connected to cardiometabolic multimorbidity.

Beyond the substantive findings, an important contribution of this study is methodological. Most mediation analyses in population health focus on a limited set of predefined mediators and outcomes. In contrast, the present framework integrates socioeconomic, psychosocial, clinical, behavioral, laboratory, and genomic information within a unified latent representation space. By combining multimodal representation learning with pathway analysis, the approach allows complex relationships to be explored without requiring investigators to specify individual mediators or disease outcomes a priori. This may be particularly valuable in large-scale biomedical datasets where important relationships are distributed across many correlated variables and data modalities.

Another notable finding was the sparsity of the pathway landscape. Although 800 candidate pathways were evaluated, only a small subset accounted for most of the observed indirect associations. Rather than supporting a diffuse pattern involving many pathways of similar magnitude, the results suggest that multimodal association structure may be concentrated within a limited number of latent dimensions. Whether similar concentration patterns occur in other disease domains and populations warrants further investigation.

Several limitations should be considered. First, the analysis was observational and cannot establish causal effects. Second, pathway estimates were calculated in latent representation space rather than on the original clinical outcome scale; consequently, effect magnitudes should be interpreted as relative measures of pathway strength rather than clinically scaled changes in disease risk. Third, latent dimensions are learned constructs that require interpretation through their observed correlates and therefore may not correspond to uniquely defined biological or psychosocial mechanisms. Fourth, complete-case integration reduced the analytic sample relative to the full All of Us cohort and may have introduced selection related to data availability. Finally, psychosocial measures were self-reported and may be subject to measurement error and reporting bias \cite{wang2024using,jiang2023causal}.

In summary, this study demonstrates how multimodal representation learning can be used to investigate pathway structure in large-scale population health data. As increasingly rich biomedical datasets become available, approaches that jointly integrate social, behavioral, clinical, and biological information may provide new opportunities to understand the complex processes linking social conditions to chronic disease. The present findings highlight psychosocial vulnerability as a prominent dimension associated with the relationship between socioeconomic disadvantage and cardiometabolic multimorbidity and motivate further investigation in longitudinal and mechanistic studies.

\section*{Conclusion}

In a multimodal cohort from the All of Us Research Program, socioeconomic disadvantage was associated with cardiometabolic multimorbidity through a latent psychosocial dimension characterized by poorer mental health, loneliness, reduced social well-being, and lower health literacy. More broadly, the study illustrates how multimodal representation learning and latent-space pathway analysis can be used to organize complex relationships within large-scale health data. Because the analysis was observational and conducted in latent representation space, the findings should be interpreted as structured indirect associations rather than evidence of causally identified mediation effects.

\section*{Funding}

The authors received no specific funding for this work.

\section*{Ethics Approval and Consent to Participate}

The All of Us Research Program is overseen by a central Institutional Review Board (IRB), the Copernicus Group IRB (Protocol Number: 20191049). The present study used de-identified participant data accessed through the All of Us Researcher Workbench under the All of Us data access and use policies. According to the All of Us Research Program data passport model, registered researchers do not require separate IRB approval from the All of Us Research Program for individual research projects conducted within the Researcher Workbench. All participants provided informed consent at enrollment in the All of Us Research Program.

\section*{Author Contributions}

C.C. performed the analyses, prepared the figures, and drafted the manuscript. C.C. and S.M. developed the study design, interpreted the results, and revised the manuscript. All authors reviewed and approved the final manuscript.

\section*{Competing Interests}

The authors declare no competing interests.

\section*{Data Availability}

Data used in this study were obtained through the National Institutes of Health All of Us Research Program Researcher Workbench using the Controlled Tier dataset (CDR version 8). Access to these data is available to approved researchers through the All of Us Research Program Researcher Workbench in accordance with program policies.

\section*{Code Availability}

Code used to generate the analyses and figures is available at:
https://github.com/congca/Causal-AI-framework-of-Environmental-and-Psychosocial-Determinants-of-Cognitive-Difficulties

\section*{Acknowledgements}

We gratefully acknowledge All of Us participants for their contributions, without whom this research would not have been possible. We also thank the National Institutes of Health's All of Us Research Program for making available the participant data examined in this study.

\bibliographystyle{apacite}
\bibliography{refs}

\section*{Supplementary Methods}

Inspection of latent loadings revealed that $Z_{T2}$ was primarily driven by socioeconomic characteristics. Higher educational attainment and household income corresponded to lower latent scores, whereas lower-income and lower-education groups exhibited substantially higher scores, indicating that $Z_{T2}$ captured socioeconomic disadvantage.

Among mediator dimensions, $Z_{M13}$ demonstrated monotonic gradients across mental health, loneliness, social well-being, and health-literacy measures. Individuals reporting poorer mental health, greater loneliness, worse social functioning, and lower confidence completing medical forms exhibited progressively higher latent scores. Accordingly, $Z_{M13}$ was interpreted as a psychosocial vulnerability factor.
To further characterize the clinical meaning of this latent mediator, we examined its associations with common chronic conditions. The strongest correlations were observed for hypertension ($r=0.385$), diabetes ($r=0.337$), and hyperlipidemia ($r=0.312$), followed by heart disease ($r=0.214$) and chronic kidney disease ($r=0.208$). These associations suggest that higher $Z_{M13}$ scores reflect a broader cardiometabolic vulnerability phenotype spanning vascular, metabolic, and cardiovascular conditions rather than a single disease process.
Within disease representations, $Z_{Y9}$ was most strongly associated with hypertension, diabetes, hyperlipidemia, obesity, chronic kidney disease, sleep apnea, and cardiovascular disease, suggesting a hypertension-centered cardiometabolic multimorbidity phenotype.

To facilitate interpretation of the dominant mediation pathway, we examined the variables most strongly associated with the key latent dimensions. The dominant exposure dimension, $Z_{T2}$, was characterized by lower income, lower educational attainment, and poorer mental health, suggesting that it primarily captured socioeconomic adversity. The dominant mediator dimension, $Z_{M13}$, was characterized by poorer mental health, greater loneliness, worse social well-being, and lower health literacy, indicating psychosocial vulnerability.

 \subsection*{Rare-Variant Gene Burden Construction}

\begin{figure}[htbp]
\centering
\includegraphics[width=0.6\textwidth]{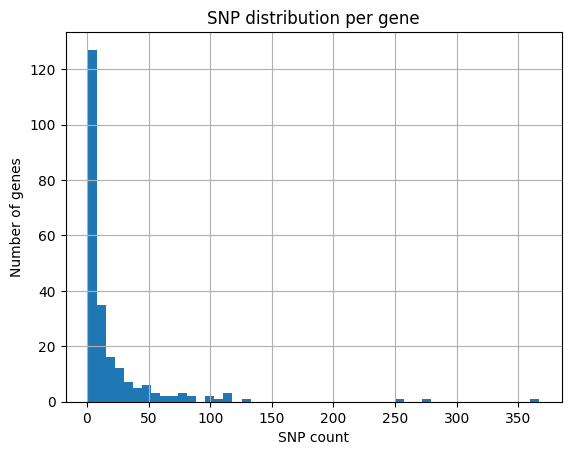}
\caption{
Distribution of rare-variant counts across genes included in burden score construction. Most genes contained relatively few observed variants, whereas a small subset exhibited substantially higher variant densities, reflecting the expected long-tailed distribution of genetic variation across the genome.
}
\label{fig:snp_distribution}
\end{figure}
 
\begin{table}[htbp]
\centering
\caption{Summary of rare-variant burden construction.}
\label{tab:burden_construction}
\begin{tabular}{lc}
\hline
Metric & Value \\
\hline
OMIM genes identified & 1,155 \\
Variants within OMIM genes & 111,331 \\
Rare variants (MAF <1\%) & 103,023 \\
Post-QC rare variants & 98,799 \\
Retained genes & 310 \\
WGS participants & 103,867 \\
Final burden features & 310 \\
\hline
\end{tabular}
\end{table}

The distribution of SNP counts per gene was highly right-skewed (Figure~\ref{fig:snp_distribution}). Most genes contained fewer than 20 SNPs, whereas a small number of genes exhibited substantially larger variant counts, producing a long-tailed distribution. This pattern is consistent with the heterogeneous genomic architecture typically observed in population-scale genetic datasets and motivates the use of representation-learning approaches that can accommodate highly imbalanced feature densities.

Mediation effects were highly concentrated within a limited subset of latent dimensions. Among exposure representations, $Z_{T2}$ exhibited the strongest overall mediation activity (mean absolute NIE = 0.000551), followed by $Z_{T4}$ (0.000331), $Z_{T3}$ (0.000327), and $Z_{T1}$ (0.000165). Among mediator dimensions, $Z_{M13}$ demonstrated the largest overall mediation contribution (0.000743), followed by $Z_{M19}$ (0.000646), $Z_{M4}$ (0.000644), $Z_{M5}$ (0.000557), and $Z_{M12}$ (0.000556). Outcome mediation effects were dominated by $Z_{Y9}$ (0.000681), followed by $Z_{Y4}$ (0.000569) and $Z_{Y8}$ (0.000515).

The strongest pathway was 
$Z_{T2} \rightarrow Z_{M13} \rightarrow Z_{Y9}$ ($\mathrm{NIE}=0.002517$). The corresponding coefficients were positive for both the exposure--mediator association ($a=0.198$) and mediator--outcome association ($b=0.0127$), indicating that greater socioeconomic disadvantage was associated with higher psychosocial vulnerability and, subsequently, greater disease burden.

\subsection*{Validation of disease-network latent representations}

To investigate the clinical meaning of the dominant disease latent factor, we evaluated correlations between $Z_{Y9}$ and routinely measured laboratory biomarkers. Among available laboratory measurements, systolic blood pressure exhibited the strongest association with $Z_{Y9}$ ($r=0.117$). In contrast, correlations with HbA1c ($r=0.020$), diastolic blood pressure ($r=0.015$), cholesterol ($r=0.002$), and glucose ($r\approx0$) were substantially weaker. These findings are consistent with the disease composition of the corresponding outcome cluster, which was enriched for hypertension, heart disease, chronic kidney disease, heart failure, atrial fibrillation, diabetes, and obesity. Taken together, the latent factor appears to represent a hypertension-centered cardiometabolic multimorbidity dimension.
These analyses were used solely for interpretation and validation of latent outcome representations and were not included in mediation estimation.

To evaluate the stability of the strongest mediation signal, we performed nonparametric bootstrap resampling for the dominant pathway

\[
Z_{T2}
\rightarrow
Z_{M13}
\rightarrow
Z_{Y9}.
\]

 \begin{table}[htbp]
\centering
\caption{
Variational autoencoder architecture and hyperparameter selection.
}
\label{tab:vae_combined}

\small

\begin{tabular}{lll}
\hline
Parameter &
Search Space &
Selected Value \\
\hline

Input features &
Fixed &
310 \\

Encoder layer 1 &
128, 256 &
256 \\

Encoder layer 2 &
128 &
128 \\

Latent dimension &
5,10,20,50 &
10 \\

Decoder layer 1 &
128 &
128 \\

Decoder layer 2 &
256 &
256 \\

Learning rate &
$10^{-3},10^{-4}$ &
$10^{-3}$ \\

KL weight &
0.001,0.01 &
0.001 \\

Batch size &
Fixed &
1024 \\

Epochs &
Fixed &
100 \\

Optimizer &
Fixed &
Adam \\

Latent distribution &
Fixed &
Gaussian \\
\hline
\end{tabular}
\end{table}

For robustness analyses, nonparametric bootstrap resampling was performed using 500 bootstrap samples. In each iteration, participants were sampled with replacement and mediation effects were re-estimated. Percentile-based 95\% confidence intervals were calculated from the empirical distribution of bootstrap estimates.

All models were implemented in PyTorch. Optimization was performed using the Adam optimizer with an initial learning rate of \(10^{-3}\). Models were trained for 100 epochs using mini-batch stochastic gradient descent with a batch size of 1,024. Data were randomly partitioned into training (80\%) and validation (20\%) subsets during hyperparameter optimization. Hyperparameter selection was based on validation-set ELBO loss. All analyses were conducted on NVIDIA GPU hardware within the All of Us Researcher Workbench environment.
\section*{Supplementary Results}

Across 500 bootstrap samples, the estimated natural indirect effect was

\[
\widehat{\mathrm{NIE}}
=
0.00291,
\]

with a percentile-based 95\% confidence interval of

\[
(0.00243,\;0.00344).
\]

The confidence interval remained entirely above zero, indicating a highly stable positive indirect effect. These results support the robustness of the identified pathway linking socioeconomic disadvantage, psychosocial vulnerability, and disease-network burden.

The average mediation magnitude of $Z_{T2}$ was approximately 3.3 times greater than that of $Z_{T1}$, highlighting the dominant contribution of socioeconomic disadvantage within the latent exposure space.

\begin{table}[htbp]
\centering
\caption{Summary of mediation architecture across exposure, mediator, and outcome latent dimensions.}
\label{tab:mediation_summary}

\small

\begin{tabular}{llccc}
\hline
Category & Dimension & Mean $|NIE|$ & Max $|NIE|$ & Relative Mean $|NIE|$ \\
\hline

\multicolumn{5}{l}{Exposure} \\
\hline
& $Z_{T2}$ & 0.000551 & 0.002517 & 1.00 \\
& $Z_{T4}$ & 0.000331 & 0.001428 & 0.60 \\
& $Z_{T3}$ & 0.000327 & 0.001518 & 0.59 \\
& $Z_{T1}$ & 0.000165 & 0.000767 & 0.30 \\
\hline

\multicolumn{5}{l}{Mediator} \\
\hline
& $Z_{M13}$ & 0.000743 & 0.002517 & 1.00 \\
& $Z_{M19}$ & 0.000646 & 0.002148 & 0.87 \\
& $Z_{M4}$  & 0.000644 & 0.002391 & 0.87 \\
& $Z_{M5}$  & 0.000557 & 0.002049 & 0.75 \\
& $Z_{M12}$ & 0.000556 & 0.001980 & 0.75 \\
& $Z_{M6}$  & 0.000551 & 0.001880 & 0.74 \\
& $Z_{M11}$ & 0.000489 & 0.001663 & 0.66 \\
& $Z_{M14}$ & 0.000381 & 0.001156 & 0.51 \\
& $Z_{M10}$ & 0.000320 & 0.001129 & 0.43 \\
& $Z_{M18}$ & 0.000307 & 0.001041 & 0.41 \\
\hline

\multicolumn{5}{l}{Outcome} \\
\hline
& $Z_{Y9}$  & 0.000681 & 0.002517 & 1.00 \\
& $Z_{Y4}$  & 0.000569 & 0.001812 & 0.84 \\
& $Z_{Y8}$  & 0.000515 & 0.001988 & 0.76 \\
& $Z_{Y10}$ & 0.000344 & 0.001071 & 0.51 \\
& $Z_{Y1}$  & 0.000286 & 0.001088 & 0.42 \\
& $Z_{Y5}$  & 0.000267 & 0.001078 & 0.39 \\
& $Z_{Y7}$  & 0.000217 & 0.000912 & 0.32 \\
& $Z_{Y2}$  & 0.000199 & 0.000666 & 0.29 \\
& $Z_{Y6}$  & 0.000186 & 0.000537 & 0.27 \\
& $Z_{Y3}$  & 0.000172 & 0.000679 & 0.25 \\
\hline
\end{tabular}

\vspace{0.2cm}

\footnotesize
Relative Mean $|NIE|$ values were normalized within each category (exposure, mediator, and outcome) to the strongest latent dimension.

\end{table}

\begin{figure}[htbp]
\centering

\includegraphics[width=.48\textwidth]{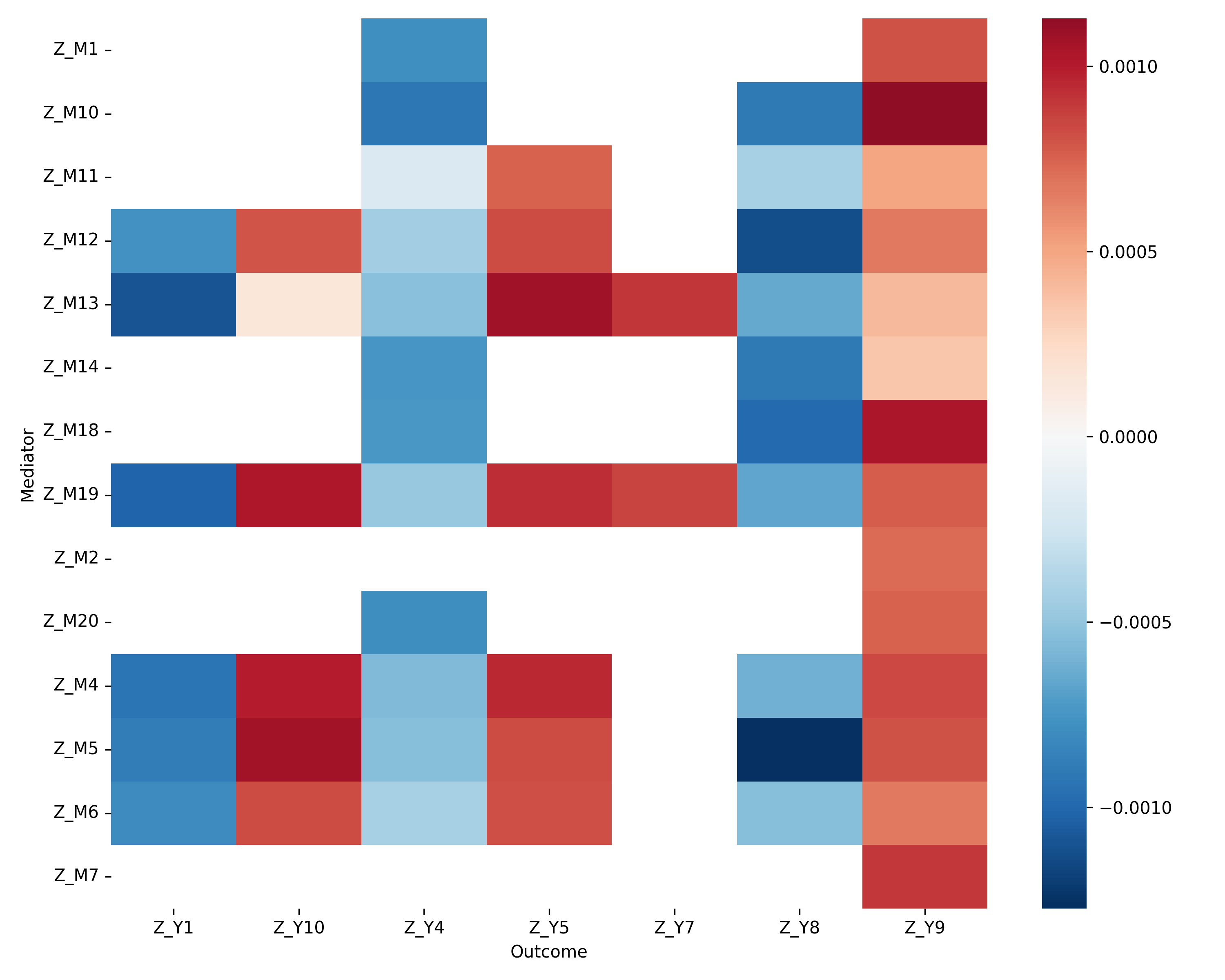}
\includegraphics[width=.48\textwidth]{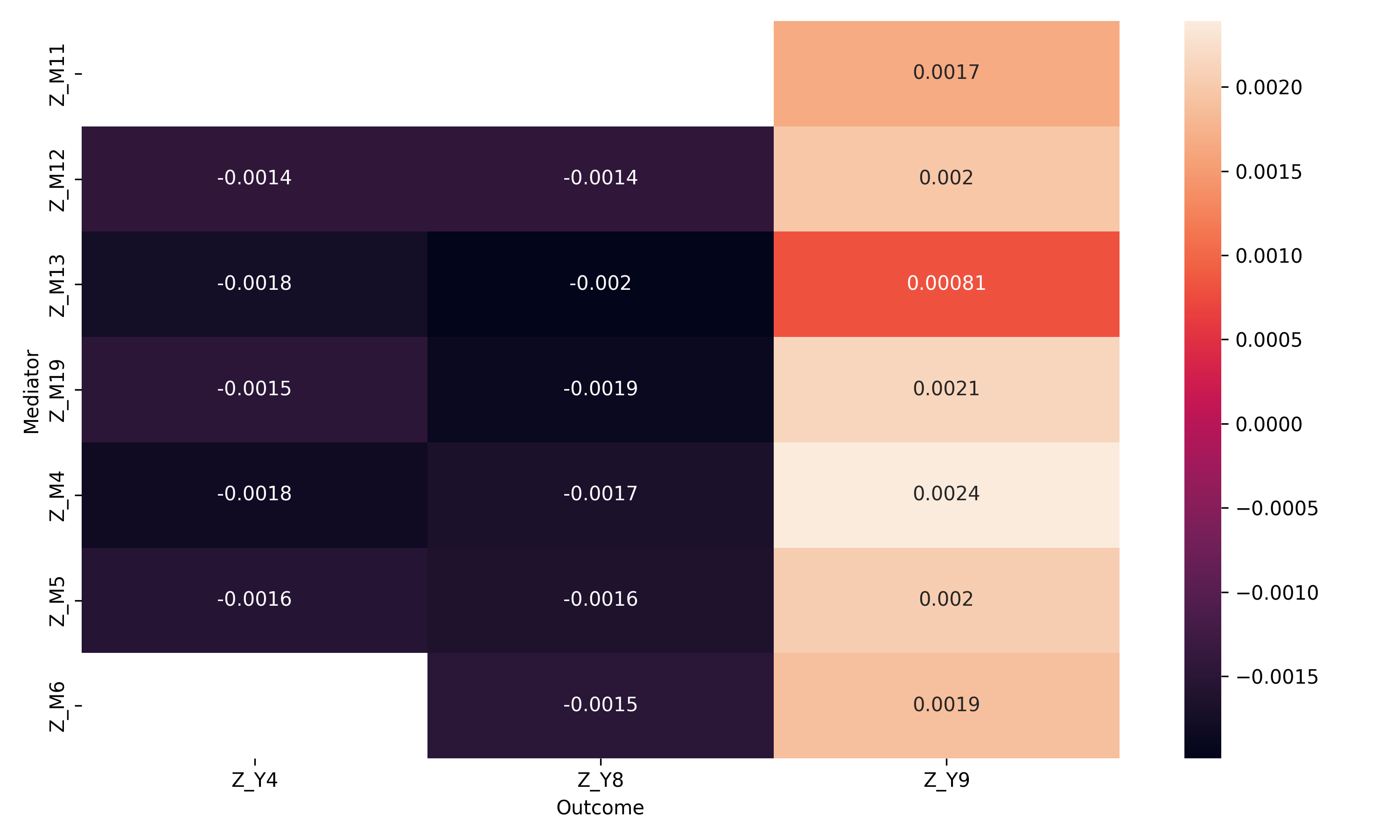}

\caption{
Supplementary mediation patterns.
(A) Mean absolute NIEs across mediator–outcome pairs.
(B) Top-ranked mediation pathways ranked by absolute NIE.
}
\end{figure}
 
  \begin{table}[htbp]
\centering
\caption{
Top-ranked latent mediation pathways and corresponding path coefficients.
}
\label{tab:top_pathways_combined}

\small

\begin{tabular}{rlllrrrr}
\hline
Rank &
Exposure &
Mediator &
Outcome &
$a$ &
$b$ &
NIE &
NDE \\
\hline

1 &
$Z_{T2}$ &
$Z_{M13}$ &
$Z_{Y9}$ &
0.198 &
0.0127 &
0.002517 &
-0.01215 \\

2 &
$Z_{T2}$ &
$Z_{M4}$ &
$Z_{Y9}$ &
-0.270 &
-0.00884 &
0.002391 &
-0.01202 \\

3 &
$Z_{T2}$ &
$Z_{M19}$ &
$Z_{Y9}$ &
-0.180 &
-0.01195 &
0.002148 &
-0.01178 \\

4 &
$Z_{T2}$ &
$Z_{M5}$ &
$Z_{Y9}$ &
-0.223 &
-0.00918 &
0.002049 &
-0.01168 \\

5 &
$Z_{T2}$ &
$Z_{M13}$ &
$Z_{Y8}$ &
0.198 &
-0.01003 &
-0.001988 &
0.01187 \\

6 &
$Z_{T2}$ &
$Z_{M12}$ &
$Z_{Y9}$ &
0.345 &
0.00574 &
0.001980 &
-0.01161 \\

7 &
$Z_{T2}$ &
$Z_{M19}$ &
$Z_{Y8}$ &
-0.180 &
0.01046 &
-0.001881 &
0.01176 \\

8 &
$Z_{T2}$ &
$Z_{M6}$ &
$Z_{Y9}$ &
0.220 &
0.00856 &
0.001880 &
-0.01151 \\

9 &
$Z_{T2}$ &
$Z_{M4}$ &
$Z_{Y4}$ &
-0.270 &
0.00670 &
-0.001812 &
0.00409 \\

10 &
$Z_{T2}$ &
$Z_{M13}$ &
$Z_{Y4}$ &
0.198 &
-0.00889 &
-0.001762 &
0.00404 \\

\hline
\end{tabular}
\end{table}

\clearpage

\small

\begin{longtable}{
p{0.09\textwidth}
p{0.15\textwidth}
p{0.17\textwidth}
p{0.40\textwidth}
p{0.12\textwidth}}

\caption{Variables included in socioeconomic, covariate, psychosocial, and laboratory modalities.}
\label{tab:all_variables}
\\

\hline
\textbf{Modality} & \textbf{Domain} & \textbf{Variable} & \textbf{Raw Answers / Measurement} & \textbf{Final Variable} \\
\hline
\endfirsthead

\hline
\textbf{Modality} & \textbf{Domain} & \textbf{Variable} & \textbf{Raw Answers / Measurement} & \textbf{Final Variable} \\
\hline
\endhead

\multicolumn{5}{l}{\textbf{Socioeconomic Exposures (T)}} \\
\hline

T & Education & Education Level & Never Attended; One Through Four; Five Through Eight; Nine Through Eleven; Twelve Or GED; College One to Three; College Graduate; Advanced Degree & Low / Medium / High \\

T & Income & Annual Income & less 10k; 10k 25k; 25k 35k; 35k 50k; 50k 75k; 75k 100k; 100k 150k; 150k 200k; more 200k & Low / Middle / High \\

T & Housing & Living Situation & Family; Friend; Motel Hotel; Shelter; Residential Facility; Transitional; Temporary Institute; Outside; College Camp & Stable vs Unstable \\

T & Housing & Home Ownership & Own; Rent; Other Arrangement & Own / Rent / Other \\

T & Housing & Household Size & 0; 1; 2; 3; 4; 5; 6; 7; 8; 9; 10; 11+ & 1 / 2--3 / 4--5 / 6+ \\

T & Housing & Housing Type & Free-standing house; apartment types; Homeless; Nursing home & Grouped housing type \\

T & Housing & Housing Conditions & Bug; Mold; Lead; Water leaks; Heat; etc. & Count index / binary indicators \\

T & Occupation & Employment Status & Employed; Self Employed; Student; Homemaker; Retired; Unable; Out of work & Employed vs Not employed \\

T & Marital & Marital Status & Married; Never; Divorced; Separated; Widowed; Partner & 3-category marital status \\

\hline
\multicolumn{5}{l}{\textbf{Baseline Covariates (X)}} \\
\hline

X & Demographic & Sex At Birth & Male; Female; Intersex; None & Male / Female / Other \\

X & Demographic & Race & AIAN; Asian; Black; Hispanic; White; etc. & White / Black / Asian / Hispanic / Other \\

X & Smoking & Smoking Frequency & Every Day; Some Days; Not At All & Smoking status \\

X & Smoking & 100 Cigarettes & Yes; No & Combine into smoking status \\

X & Alcohol & Alcohol Participation & Yes; No & Drink / No drink \\

X & Alcohol & Frequency & Never; Monthly; Weekly; Daily & Low / Moderate / High \\

X & Alcohol & Quantity & 1--2; 3--4; 5--6; etc. & Combined drinking level \\

X & Drugs & Marijuana Use & Daily; Weekly; Monthly; Once; Never & Any use vs none \\

X & Drugs & Cocaine Use & Daily; Weekly; Monthly; Once; Never & Any use vs none \\

X & Drugs & Opioid Use & Daily; Weekly; Monthly; Once; Never & Any use vs none \\

X & Drugs & Drug Type & Multiple categories & Binary indicators \\

X & Family History & Family Disease & Self; Parent; Sibling; etc. & Optional covariate \\

\hline
\multicolumn{5}{l}{\textbf{Psychosocial Mediators (M)}} \\
\hline

M & Mental Health & General Mental Health & Excellent; Very Good; Good; Fair; Poor & Ordinal score \\

M & Emotional & Emotional Distress & Always; Often; Sometimes; Rarely; Never & Low / Medium / High \\

M & Social & Social Satisfaction & Excellent; Very Good; Good; Fair; Poor & Ordinal \\

M & Literacy & Medical Form Confidence & Not at all $\rightarrow$ Extremely & Ordinal \\

M & Literacy & Understanding Info & Always $\rightarrow$ Never & Ordinal \\

M & Support & Material Assistance & Always $\rightarrow$ Never & Ordinal \\

M & Loneliness & Social Frequency & Multiple frequency levels & Low / Medium / High \\

\hline
\multicolumn{5}{l}{\textbf{Laboratory Covariates (L)}} \\
\hline

L & Anthropometric & Body mass index (BMI) [Ratio] & LOINC 39156-5 & BMI (continuous) \\

L & Vital Signs & Systolic blood pressure & LOINC 8480-6 & Blood pressure \\

L & Metabolic & Glucose [Mass/volume] in Serum or Plasma & LOINC 2345-7 & Glucose / Diabetes marker \\

L & Renal Function & Creatinine [Mass/volume] in Serum or Plasma & LOINC 2160-0 & Renal function \\

L & Hematology & Hemoglobin [Mass/volume] in Blood & LOINC 718-7 & Hemoglobin \\

L & Hematology & Leukocytes [/volume] in Blood & LOINC 6690-2 & WBC count \\

L & Lipids & Cholesterol [Mass/volume] in Serum or Plasma & LOINC 2093-3 & Lipid panel \\

L & Lipids & Cholesterol in HDL [Mass/volume] & LOINC 2085-9 & Lipid profile (HDL) \\

L & Liver Function & Alanine aminotransferase (ALT) & LOINC 1742-6 & Liver function \\

L & Endocrine & Thyrotropin (TSH) & LOINC 3016-3 & Thyroid function \\

\hline
\end{longtable}

\subsection*{Disease Network Construction and Included Diseases}
We curated a set of common diseases representing high-prevalence conditions 
in U.S. populations based on prior epidemiological reports and literature, 
including sources from the Centers for Disease Control and Prevention (CDC) 
and the Peterson-KFF Health System Tracker.
\cite{bechman2025incidence}
\cite{garnett2026overdose}
\cite{telesford2025chronic}

The disease list focuses on clinically well-defined conditions across major organ systems, including cardiovascular, metabolic, respiratory, and mental health,  musculoskeletal, digestive, renal, neurological, ophthalmologic, dermatologic,  and infectious diseases. Non-specific or overly broad categories were excluded to ensure interpretability in downstream network analysis.

\small

\begin{table}[htbp]
\centering
\small
\caption{Disease Categories and Conditions Used for Analysis}
\begin{tabular}{ll}
\hline
\textbf{Category} & \textbf{Disease} \\
\hline

General & Stroke \\
General & Lung Disease \\
General & Cancer \\
General & High Blood Pressure \\
General & Diabetes \\
General & Heart Disease \\
General & Heart Attack \\

Cardiovascular \& Metabolic & Hypertension \\
Cardiovascular \& Metabolic & Type 2 Diabetes \\
Cardiovascular \& Metabolic & Hyperlipidemia \\
Cardiovascular \& Metabolic & Coronary Artery Disease \\
Cardiovascular \& Metabolic & Heart Failure \\
Cardiovascular \& Metabolic & Atrial Fibrillation \\
Cardiovascular \& Metabolic & Stroke \\
Cardiovascular \& Metabolic & Peripheral Arterial Disease \\

Metabolic \& Endocrine & Obesity \\
Metabolic \& Endocrine & Metabolic Syndrome \\
Metabolic \& Endocrine & Hypothyroidism \\
Metabolic \& Endocrine & Hyperthyroidism \\

Respiratory & Asthma \\
Respiratory & Chronic Obstructive Pulmonary Disease \\
Respiratory & Sleep Apnea \\

Mental Health & Depression \\
Mental Health & Anxiety Disorder \\
Mental Health & Substance Use Disorder \\
Mental Health & Bipolar Disorder \\

Musculoskeletal & Osteoarthritis \\
Musculoskeletal & Rheumatoid Arthritis \\
Musculoskeletal & Osteoporosis \\

Digestive & Gastroesophageal Reflux Disease \\
Digestive & Irritable Bowel Syndrome \\
Digestive & Fatty Liver Disease \\
Digestive & Chronic Liver Disease \\

Kidney \& Hematologic & Chronic Kidney Disease \\
Kidney \& Hematologic & Anemia \\

Infectious \& Other & Urinary Tract Infection \\
Infectious \& Other & Sepsis \\

Ophthalmologic \& Neurologic & Cataract \\
Ophthalmologic \& Neurologic & Glaucoma \\
Ophthalmologic \& Neurologic & Migraine \\

Dermatologic & Dermatitis \\
Dermatologic & Psoriasis \\

\hline
\end{tabular}
\end{table}

Gene–disease associations were obtained from the OMIM database, 
which provides high-confidence curated associations, and further 
supplemented using curated data from the DisGeNET database to improve 
coverage of common complex diseases.

\end{document}